\documentclass[conference,usenames,table,xcdraw]{IEEEtran}

\usepackage{import}
\usepackage[numbers,compress]{natbib}

\usepackage{mathptmx}   
\usepackage{emptypage}  
\usepackage{blindtext}  
\usepackage{ragged2e}   
\usepackage{pdflscape}  
\usepackage{multicol}   
\usepackage{gensymb}    
\usepackage{bold-extra} 
\usepackage{amsmath}    
\usepackage{amssymb}    
\usepackage{hyperref}   
\usepackage{cleveref}   

\makeatletter
\let\MYcaption\@makecaption
\makeatother

\usepackage[font=footnotesize]{subcaption}

\makeatletter
\let\@makecaption\MYcaption
\makeatother

\let\labelindent\relax
\usepackage{enumitem}
\usepackage[multiple]{footmisc} 

\usepackage{graphicx}
\graphicspath{{images/}{../images/}}

\usepackage{listings}
\usepackage{textcomp}
\usepackage{color}
\definecolor{maroon}{rgb}{0.5,0,0}
\definecolor{darkgreen}{rgb}{0,0.5,0}

\newcommand{\beginlstdelim}[3]
{%
  \def\endlstdelim{#2\egroup}%
  \ttfamily#1\bgroup\color{#3}\aftergroup\endlstdelim%
}

\lstdefinestyle{xml_style}
{
  morestring=[b]",
  moredelim=*[s][\bfseries\color{maroon}]{<}{>},
  moredelim=*[s][\bfseries\color{maroon}]{</}{>},
  morecomment=[s]{<?}{?>},
  morecomment=[s]{<!--}{-->},
}

\lstdefinestyle{json_style}
{
  morestring=[b]",
  comment=[l]{//},
  keywordstyle=\bfseries\color{maroon}
}

\lstdefinelanguage{JSON}
{
  style=json_style
}

\lstdefinelanguage{XML}
{
  style=xml_style
}

\usepackage{booktabs}
\usepackage{multirow}
\usepackage{makecell}

\usepackage{xcolor}

\title{Losing Confidence in Quality:\\Unspoken Evolution of Computer Vision Services}

\author{

\IEEEauthorblockN{
  Alex Cummaudo\IEEEauthorrefmark{1},
  Rajesh Vasa\IEEEauthorrefmark{1},
  John Grundy\IEEEauthorrefmark{2},
  Mohamed Abdelrazek\IEEEauthorrefmark{3} and
  Andrew Cain\IEEEauthorrefmark{3}
} 
\IEEEauthorblockA{
  \IEEEauthorrefmark{1}
  Applied Artificial Intelligence Institute, Deakin University, Geelong, Australia\\
} 
\IEEEauthorblockA{
  \IEEEauthorrefmark{2}
  Faculty of Information Technology, Monash University, Clayton, Australia\\
}
\IEEEauthorblockA{
  \IEEEauthorrefmark{3}
  School of Information Technology, Deakin University, Geelong, Australia\\
  {\itshape \{ ca, rajesh.vasa, mohamed.abdelrazek, andrew.cain \}@deakin.edu.au, john.grundy@monash.edu}
}
}

\usepackage{balance}

\usepackage{siunitx}
\usepackage{multibib}
\newcites{web}{Online Artefacts} 

\def \googleapi {A}
\def \azureapi {B}
\def \awsapi {C}

\begin{document}

\maketitle
\begin{abstract}
Recent advances in artificial intelligence (AI) and machine learning (ML), such as computer vision, are now available as intelligent services and their accessibility and simplicity is compelling. Multiple vendors now offer this technology as cloud services and developers want to leverage these advances to provide value to end-users. 
However, there is no firm investigation into the maintenance and evolution risks arising from use of these intelligent services; in particular, their behavioural consistency and transparency of their functionality.
We evaluated the responses of three different intelligent services (specifically  computer vision) over 11 months using 3 different data sets, verifying responses against the respective documentation and assessing evolution risk. 
We found that there are: (1) inconsistencies in how these services behave; (2) evolution risk in the responses; and (3) a lack of clear communication that documents these risks and inconsistencies.
We propose a set of recommendations to both developers and intelligent service providers to inform risk and assist maintainability.
\end{abstract}

\IEEEpeerreviewmaketitle

\section{Introduction}
The availability of intelligent services has made AI tooling accessible to software developers and promises a lower entry barrier for their utilisation. Consider state-of-the-art computer vision analysers, which require either manually training a deep-learning classifier, or selecting a pre-trained model and deploying these into an appropriate infrastructure. Either are laborious in time, and require non-trivial expertise along with a large data set when training or customisation is needed.
In contrast, intelligent services providing computer vision (e.g.,~\citepweb{GoogleCloud:Home,Azure:Home,AWS:Home,Pixlab:Home,IBM:Home,Cloudsight:Home,Clarifai:Home,DeepAI:Home,Imagaa:Home,Talkwaler:Home,Megvii:Home,TupuTech:Home,YiTuTech:Home,SenseTime:Home}) abstract these complexities behind a web API call. This removes the need to understand the complexities required of ML, and requires little more than the knowledge on how to use RESTful endpoints. The ubiquity of these services is exemplified through their rapid uptake in applications such as aiding the vision-impaired~\citep{Reis:2018cp,daMotaSilveira:2017vp}.

While intelligent services have seen quick adoption in industry, there has been little work that has considered the software quality perspective of the risks and impacts posed by using such services. In relation to this, there are three main challenges: (1) incorporating stochastic algorithms into software that has traditionally been deterministic; (2) the general lack of transparency associated with the ML models; and (3) communicating to application developers.

ML typically involves use of statistical techniques that yield components with a non-deterministic external behaviour; that is, for the same given input, different outcomes may result. However, developers, in general, are used to libraries and small components behaving predictably, while systems that rely on ML techniques work on confidence intervals\footnote{Varied terminology used here. Probability, confidence, accuracy and score may all be used interchangeably.} and probabilities. For example, the developer's mindset suggests that an image of a border collie---if sent to three intelligent computer vision services---would return the label `dog' consistently with time regardless of which service is used. However, one service may yield the specific dog breed, `border collie', another service may yield a permutation of that breed, `collie', and another may yield broader results, such as `animal'; each with results of varying confidence values.\footnote{Indeed, we have observed this phenomenon using a picture of a border collie sent to various computer vision services.}  
Furthermore, the third service may evolve with time, and thus learn that the `animal' is actually a `dog' or even a `collie'. The outcomes are thus behaviourally inconsistent between services providing conceptually similar functionality.
As a thought exercise, consider if the sub-string function were created using ML techniques---it would perform its operation with a confidence where the expected outcome and the AI inferred output match as a \textit{probability}, rather than a deterministic (constant) outcome. How would this affect the developers' approach to using such a  function? Would they actively take into consideration the non-deterministic nature of the result? 

Myriad software quality models and SE practices advocate maintainability and reliability as primary characteristics; stability, testability, fault tolerance, changeability and maturity are all concerns for quality in software components~\citep{Pressman:2005vf,Sommerville:2011uc,Horch:2003uv} and one must factor these in with consideration to software evolution challenges~\citep{demeyer2008software,4659256,tu2000evolution,1572302,THOMAS2014457}. However, the effect this non-deterministic behaviour has on quality when masked behind  an intelligent service is still under-explored to date in SE literature, to our knowledge. Where software depends on intelligent services to achieve functionality, these quality characteristics may not be achieved, and developers need to be wary of the unintended side effects and inconsistency that exists when using non-deterministic components. A computer vision service may encapsulate deep-learning strategies or stochastic methods to perform image analysis, but developers are more likely to approach intelligent services with a mindset that anticipates consistency. Although the documentation does hint at this non-deterministic behaviour (i.e., the descriptions of `confidence' in various computer vision services suggest the they are not always confident, and thus not deterministic  \citepweb{Google:DocsLabel,AWS:DetectLabel,Azure:HowToCall}), the integration mechanisms offered by popular vendors do not seem to fully expose the nuances, and developers are not yet familiar with the trade-offs.

Do popular computer vision services, as they currently stand, offer consistent behaviour, and if not, how is this conveyed to developers (if it is at all)? If computer vision services are to be used in production services, do they ensure quality under rigorous Software Quality Assurance (SQA) frameworks~\citep{Horch:2003uv}? What evolution risk~\citep{demeyer2008software,4659256,tu2000evolution,1572302} do they pose if these services change?
To our knowledge, few studies have been conducted to investigate these claims. This paper assesses the consistency, evolution risk and consequent maintenance issues that may arise when developers use intelligent services. We introduce a motivating example in \cref{sec:motivating-example}, discussing related work and our methodology in \cref{sec:related-work,sec:method}. We present and interpret our findings in \cref{sec:findings}. We argue with quantified evidence that these intelligent services can only be considered with a mature appreciation of risks, and we make a set of recommendations in \cref{sec:recommendations}.

\section{Motivating Example}
\label{sec:motivating-example}

Consider Rosa, a software developer, who wants to develop a social media photo-sharing mobile app that analyses her and her friends photos on Android and iOS. Rosa wants the app to categorise photos into scenes (e.g., day vs. night, outdoors vs. indoors), generate brief descriptions of each photo, and catalogue photos of her friends as well as common objects (e.g., all photos with a dog, all photos on the beach).

Rather than building a computer vision engine from scratch, Rosa thinks she can achieve this using one of the popular computer vision services (e.g.,~\citepweb{GoogleCloud:Home,Azure:Home,AWS:Home,Pixlab:Home,IBM:Home,Cloudsight:Home,Clarifai:Home,DeepAI:Home,Imagaa:Home,Talkwaler:Home,Megvii:Home,TupuTech:Home,YiTuTech:Home,SenseTime:Home}). However, Rosa comes from a typical software engineering background with limited knowledge of the underlying deep-learning techniques and implementations as currently used in computer vision. Not unexpectedly, she internalises a mindset of how such services work and behave based on her experience of using software libraries offered by various SDKs. This mindset assumes that different cloud vendor image processing APIs more-or-less provide similar functionality, with only minor variations. For example, cloud object storage for Amazon S3 is both conceptually and behaviourally very similar to that of Google Cloud Storage or Azure Storage. Rosa assumes the computer vision services of these platforms will, therefore, likely be very similar. Similarly, consider the string libraries Rosa will use for the app. The conceptual and behavioural similarities are consistent; a string library in Java (Android) is conceptually very similar to the string library she will use in Swift (iOS), and likewise both behave similarly by providing the same results for their respective sub-string functionality. However, \textbf{unlike the cloud storage and string libraries, different computer vision services often present conceptually similar functionality but are behaviourally very different}. Intelligent service vendors also hide the depth of knowledge needed to use these effectively---for instance, the training data set and ontologies used to create these services are hidden in the documentation. Thus, Rosa isn't even exposed to this knowledge as she reads through the documentation of the providers and, thus, Rosa makes the following assumptions:

\begin{itemize}
  \item \textbf{``I think the responses will be consistent amongst these computer vision services.''} When Rosa uploads a photo of a dog, she would expect them all to respond with `dog'. If Rosa decides to switch which service she is using, she expects the ontologies to be compatible (all computer vision services \textit{surely} return dog for the same image) and therefore she can expect to plug-in a different service should she feel like it making only minor code modifications such as which endpoints she is relying on.
  \item \textbf{``I think the responses will be constant with time.''} When Rosa uploads the photo of a dog for testing, she expects the response to be the same in 10 weeks time once her app is in production. Hence, in 10 weeks, the same photo of the dog should return the same label.
\end{itemize}

\section{Related Work}
\label{sec:related-work}

If we were to view computer vision services through the lenses of an SQA framework, robustness, consistency, and maintainability often feature as quality attributes in myriad software quality models (e.g., \citep{ISO9126:1999}). Software quality is determined from two key dimensions: (1) in the evaluation of the end-product (external quality) and (2) the assurances in the development processes (internal quality)~\citep{Pressman:2005vf}.  We discuss both perspectives of quality within the context of our work in this section.

\subsection{External Quality}

\subsubsection{Robustness for safety-critical applications}
A typical focus of recent work has been to investigate the robustness of deep-learning within computer vision technique implementation, thereby informing the effectiveness in the context of the end-product. The common method for this has been via the use of adversarial examples~\citep{Szegedy:2013vw}, where input images are slightly perturbed to maximise prediction error but are still interpretable to humans.

Google Cloud Vision, for instance, fails to correctly classify adversarial examples when noise is added to the original images~\citep{Hosseini:2018jr}. \citet{Rosenfeld:2018ut} illustrated that inserting synthetic foreign objects to input images (e.g., a cartoon elephant) can completely alter classification output. \citet{Wang:2018vl} performed similar attacks on a transfer-learning approach of facial recognition by modifying pixels of a celebrity's face to be recognised as a completely different celebrity, all while still retaining the same human-interpretable original celebrity. \citet{Su:2017uw} used the ImageNet database to show that 41.22\% of images drop in confidence when just a \textit{single pixel} is changed in the input image; and similarly, \citet{Eykholt:2018vka} recently showed similar results that made a CNN interpret a stop road-sign (with mimicked graffiti) as a 45mph speed limit sign.

The results suggest that current state-of-the-art computer vision techniques may not be robust enough for safety critical applications as they do not handle intentional or unintentional adversarial attacks. Moreover, as such adversarial examples exist in the physical world~\citep{Kurakin:2016vw,Eykholt:2018vk}, ``the natural world may be adversarial enough''~\citep{Pezzementi:2018tq} to fool AI software. Though some limitations and guidelines have been explored in this area, the perspective of \textit{intelligent} services is yet to be considered and specific guidelines do not yet exist when using computer vision services.

\subsubsection{Testing strategies in ML applications}

Although much work applies ML techniques to automate testing strategies, there is only a growing emphasis that considers this in the opposite sense; that is, testing to ensure the ML product works correctly. There are few reliable test oracles that ensure if an ML has been implemented to serve its algorithm and use case purposefully; indeed, ``the non-deterministic nature of many training algorithms makes testing of models even more challenging'' \citep{Arpteg:2018ke}.
\citet{murphy2007approach} proposed a SE-based testing approach on ML ranking algorithms to evaluate the `correctness' of the implementation on a real-world data set and problem domain, whereby discrepancies were found from the formal mathematical proofs of the ML algorithm and the implementation. 

Recently, \citet{braiek2018testing} conducted a comprehensive review of testing strategies in ML software, proposing several research directions and recommendations in how best to apply SE testing practices in ML programs. However, much of the area of this work specifically targets ML engineers, and not application developers. Little has been investigated on how application developers perceive and understand ML concepts, given a lack of formal training; we note that other testing strategies and frameworks proposed (e.g., \citep{nishi2018test,murphy2008improving,breck2016s}) are targeted chiefly to the ML engineer, and not the application developer.

However, \citet{Arpteg:2018ke} recently demonstrated (using real-world ML projects) the developmental challenges posed to developers, particularly those that arise when there is a lack of transparency on the models used and how to troubleshoot ML frameworks using traditional SE debugging tools. This said, there is no further investigations into challenges when using the higher, `ML friendly' layers (e.g., intelligent services) of the `machine learning spectrum' \citep{Ortiz:2017wg}, rather than the `lower layers' consisting of existing ML frameworks and algorithms targeted toward the ML community.

\subsection{Internal Quality}


 
\subsubsection{Quality metrics for cloud services}
 
Computer vision services are based on cloud computing fundamentals under a subset of the Platform as a Service (PaaS) model.  There has been work in the evaluation of PaaS in terms of quality attributes~\citep{Garg:2011gw}: these attributes are exposed using Service Level Agreements (SLA) between vendors and customers, and customers denote their demanded Quality of Service (QoS) to ensure the cloud services adhere to measurable KPI attributes. 

Although, popular services, such as cloud object storage, come with strong QoS agreement, to date intelligent services do not come with deep assurances around their performance and responses, but do offer uptime guarantees. For example, how can Rosa demand a QoS that ensures all photos of dogs uploaded to her app guarantee the specific dog breeds are returned so that users can look up their other friend's `border collie's? If dog breeds are returned, what ontologies exist for breeds? Are they consistent with each other, or shortened? (`Collie' versus `border collie'; `staffy' versus `staffordshire bull terrier'?) For some applications, these unstated QoS metrics specific to the ML service may have significant legal ramifications.

\subsubsection{Web service documentation and documenting ML}

From the \textit{developer's} perspective, little has been achieved to assess intelligent service quality or assure quality of these computer vision services. Web services and their interfaces (APIs) are the bridge between developers' needs and the software components~\citep{Arnold:2005vc}; therefore, assessing such computer vision services from the quality of their APIs is thereby directly related to the development quality~\citep{Ko:2004td}. Good APIs should be intuitive and require less documentation browsing~\citep{Piccioni:2013em}, thereby increasing productivity. Conversely, poor APIs that are hard to understand and work with reduce developer productivity, thereby reducing product quality. This typically leads to developers congregating on forums such as Stack Overflow, leading to a repository of unstructured knowledge likely to concern API design \citep{7180082}. The consequences of addressing these concerns in development leads to a higher demand in technical support (as measured in~\citep{Henning:2009hz}) that, ultimately, causes the maintenance to be far more expensive, a phenomenon widely known in software engineering economics~\citep{Boehm:1981ua}. Rosa, for instance, isn't aware of technical ML concepts; if she cannot reason about what search results are relevant when browsing the service and understanding functionality, her productivity is significantly decreased. Conceptual understanding is critical for using APIs, as demonstrated by \citeauthor{Ko:2011fb}, and the effects of maintenance this may have in the future of her application is unknown.

Recent attempts to document attributes and characteristics on ML models have been proposed. Model cards were introduced by~\citet{Mitchell:2018in} to describe how particular models were trained and benchmarked, thereby assisting users to reason if the model is right for their purposes and if it can achieve its stated outcomes. \citet{Gebru:2018wh} also proposed datasheets, a standardised documentation format to describe the need for a particular data set, the information contained within it and what scenarios it should be used for, including legal or ethical concerns. 

However, while target audiences for these documents may be of a more technical AI level (i.e., the ML engineer), there is still no standardised communication format for application developers to reason about using particular intelligent services, and the ramifications this may have on the applications they write is not fully conveyed. Hence, our work is focused on the application developer perspective. 

\section{Method}
\label{sec:method}

This study organically evolved by observing phenomena surrounding computer vision services by assessing both their documentation and responses. We adopted a mixed methods approach, performing both qualitative and quantitative data collection on these two key aspects by using documentary research methods for inspecting the documentation and structured observations to quantitatively analyse the results over time. This, ultimately, helped us shape the following research hypotheses which this paper addresses:

\smallskip
\begin{enumerate}[label=\textbf{[RH\arabic*]}, leftmargin=4\parindent]
\item Computer vision services do not respond with consistent outputs between services, given the same input image.
\item The responses from computer vision services are non-deterministic and evolving, and the same service can change its top-most response over time given the same input image.
\item Computer vision services do not effectively communicate this evolution and instability, introducing risk into engineering these systems.
\end{enumerate}
\smallskip

We conducted two experiments to address these hypotheses against three popular computer vision services: AWS Rekognition~\citepweb{AWS:Home}, Google Cloud Vision~\citepweb{GoogleCloud:Home}, Azure Computer Vision~\citepweb{Azure:Home}. Specifically, we targeted the AWS \texttt{DetectLabels} endpoint~\citepweb{AWS:DetectLabel}, the Google Cloud Vision \texttt{annotate:images} endpoint~\citepweb{Google:DocsLabel} and Azure's \texttt{analyze} endpoint~\citepweb{Azure:HowToCall}. For the remainder of this paper, we de-identify our selected computer vision services by labelling them as services A, B and C but do not reveal mapping to prevent any implicit bias.
Our selection criteria for using these particular three services are based on the weight behind each service provider given their prominence in the industry (Amazon, Google and Microsoft), the ubiquity of their hosting cloud platforms as industry leaders of cloud computing (i.e., AWS, Google Cloud and Azure), being in the top three most adopted cloud vendors in enterprise applications in 2018~\citep{RightScaleInc:2018kJ} and the consistent popularity of discussion amongst developers in developer communities such as Stack Overflow. While we choose these particular cloud computer vision services, we  acknowledge that similar services~\citepweb{IBM:Home,Pixlab:Home,Clarifai:Home,Cloudsight:Home,DeepAI:Home,Imagaa:Home,Talkwaler:Home} also exist, including other popular services used in Asia~\citepweb{Megvii:Home,TupuTech:Home,YiTuTech:Home,SenseTime:Home} (some offering 3D image analysis~\citepweb{DeepGlint:Home}). We reflect on the impacts this has to our study design in \cref{sec:limitations}.

Our study involved an 11-month longitudinal study which consisted of two 13 week and 17 week experiments from April to August 2018 and November 2018 to March 2019, respectively. Our investigation into documentation occurred on August 28 2018. In total, we assessed the services with three data sets; we first ran a pilot study using a smaller pool of 30 images to confirm the end-points remain stable, re-running the study with a larger pool of images of 1,650 and 5,000 images. Our selection criteria for these three data sets were that the images had to have varying objects, taken in various scenes and various times. Images also needed to contain disparate objects. Our small data set was sourced by the first author by taking photos of random scenes in an afternoon, whilst our second data set was sourced from various members of our research group from their personal photo libraries. We also wanted to include a data set that was publicly available prior to running our study, so for this data set we chose the COCO 2017 validation data set~\citep{Lin:2014vma}. We have made our other two data sets available online (\citepweb{Dataset:Large}). We collected results and their responses from each service's API endpoint  using a  python script~\citepweb{Tooling:Scraper} that sent requests to each service periodically via cron jobs. \Cref{tab:dataset} summarises various characteristics about the data sets used in these experiments.

\begin{table}[t]
\renewcommand{\arraystretch}{1.3}
\caption{Characteristics of our data sets and responses.}
\label{tab:dataset}
\centering
\begin{tabular}{|c||c|c|c|}
  \hline
  \textbf{Data set} & \textbf{Small} & \textbf{Large} & \textbf{COCOVal17} \\
  \hline\hline
  \# Images/data set & 30 & 1,650 & 5000 \\
  \hline
  \# Unique labels found & 307 & 3506 & 4507\\
  \hline
  Number of snapshots & 9 & 22 & 22 \\
  \hline
  Avg. days b/n requests & 12 Days & 8 Days & 8 Days \\
  \hline
\end{tabular}
\end{table}


We then performed quantitative analyses on each response's labels, ensuring all labels were lowercased as case changed for services \googleapi{} and \awsapi{} over the evaluation period. To derive at the consistency of responses for each image, we considered only the `top' labels per image for each service and data set. That is, for the same image $i$ over all images in data set $D$ where $i \in D$ and over the three services, the top labels per image ($T_{i}$) of all labels per image $L_{i}$ (i.e., $T_{i} \subseteq L_{i}$) is that where the respective label's confidences are consistently the highest of all labels returned. 
Typically, the top labels returned is a set containing only one element---that is, only one unique label consistently returned with the highest label ($| T_{i} | = 1$)---however there are cases where the top labels contains multiple elements as their respective confidences are \textit{equal} ($| T_{i} | > 1$).

We measure response consistency under 6 aspects:

\begin{enumerate}[label=\textbf{\arabic*)}]
  \item \textbf{Consistency of the top label between each service.} Where the same image of, for example, a dog is sent to the three services, the top label for service A may be `animal', B `canine' and C `animal'. Therefore, service B is inconsistent.
  \item \textbf{Semantic consistency of the top labels.} Where a service has returned multiple top labels ($| T_{i} | > 1$), there may lie semantic differences in what the service thinks the image best represents. Therefore, there is conceptual inconsistency in the top labels for a service even when the confidences are equal.
  \item \textbf{Consistency of the top label's confidence per service.} The top label for an image does not guarantee a high confidence. Therefore, there may be inconsistencies in how confident the top labels for all images in a service is.\bigskip
  \item \textbf{Consistency of confidence in the intersecting top label between each service.} The spread of a top intersecting label, e.g. `cat', may not have the same confidences per service even when all three services agree that `cat' is the top label. Therefore, there is inconsistency in the confidences of a top label even where all three services agree.
  \item \textbf{Consistency of the top label over time.} Given an image, the top label in one week may differ from the top label the following week. Therefore, there is inconsistency in the top label itself due to model evolution.
  \item \textbf{Consistency of the top label's confidence over time.} The top label of an image may remain static from one week to the next for the same service, but its confidence values may change with time. Therefore, there is inconsistency in the top label's confidence due to model evolution.
\end{enumerate}

For the above aspects of consistency, we calculated the spread of variation for the top label's confidences of each service for every 1 percent point; that is, the frequency of top label confidences within 100--99\%, 99--98\% etc. The consistency of top label's and their confidences between each service was determined by intersecting the labels of each service per image and grouping the intersecting label's confidences together. This allowed us to determine relevant probability distributions. For reproducibility, all quantitative analysis is available online~\citepweb{Results:AnalysisCalculation}. 

\section{Findings}
\label{sec:findings}

\subsection{Consistency of top labels}
\label{ssec:findings:consistency-of-labels}

\subsubsection{Consistency across services}

\begin{figure}[t]
  \centering
  \includegraphics[width=0.55\linewidth]{images/000000009590}    
  \caption{The only consistent label for the above image is `people' for services \awsapi{} and \azureapi{}. The top label for \googleapi{} is `conversation' and this label is not registered amongst the other two services.}
  \label{fig:sample-images:people}
\end{figure}

\begin{table}[t]
\renewcommand{\arraystretch}{1.3}
\caption{Ratio of the top labels (to images) that intersect in each data set for each permutation of service.}
\label{tab:intersect-of-labels}
\centering
\begin{tabular}{|c||c|c|c||c|c|}
  \hline
  \textbf{Service} & \textbf{Small} & \textbf{Large} & \textbf{C'Val17} & \textbf{$\mu$} & \textbf{$\sigma$} \\
  \hline
  \hline
  \googleapi{} $\cap$ \azureapi{} $\cap$ \awsapi{}  & 3.33\%   & 2.73\%   & 4.68\%   & 2.75\%  & 0.0100  \\
  \hline
  \googleapi{} $\cap$ \azureapi{}                   & 6.67\%   & 11.27\%  & 12.26\%  & 10.07\% & 0.0299 \\
  \hline
  \googleapi{} $\cap$ \awsapi{}                     & 20.00\%  & 13.94\%  & 17.28\%  & 17.07\% & 0.0304 \\
  \hline
  \azureapi{} $\cap$ \awsapi{}                      & 6.67\%   & 12.97\%  & 20.90\%  & 13.51\% & 0.0713 \\
  \hline
\end{tabular}
\end{table}

\Cref{tab:intersect-of-labels} presents the consistency of the top labels between data sets, as measured by the cardinality of the intersection of all three services' set of top labels divided by the number of images per data set. A combination of services present varied overlaps in their top labels; services \googleapi{} and \awsapi{} provide the best overlap for all three data sets, however the intersection of all three irrespective of data sets is low.

The implication here is that, without semantic comparison (see \cref{sec:limitations}), service vendors are not `plug-and-play'. If Rosa uploaded the sample images in this paper to her application to all services, she would find that only \cref{fig:sample-images:people} responds with `person' for services  \azureapi{} and \awsapi{} in their respective set of top labels. However, if she decides to then adopt service \googleapi{}, then \cref{fig:sample-images:people}'s top label becomes `conversation'; the `person' label does not appear within the top 15 labels for service \googleapi{} and, conversely, the `conversation' label does not appear in the other services top 15. 

\begin{figure}[t]
  \centering
  \begin{subfigure}[b]{0.49\linewidth}
    \includegraphics[width=\linewidth]{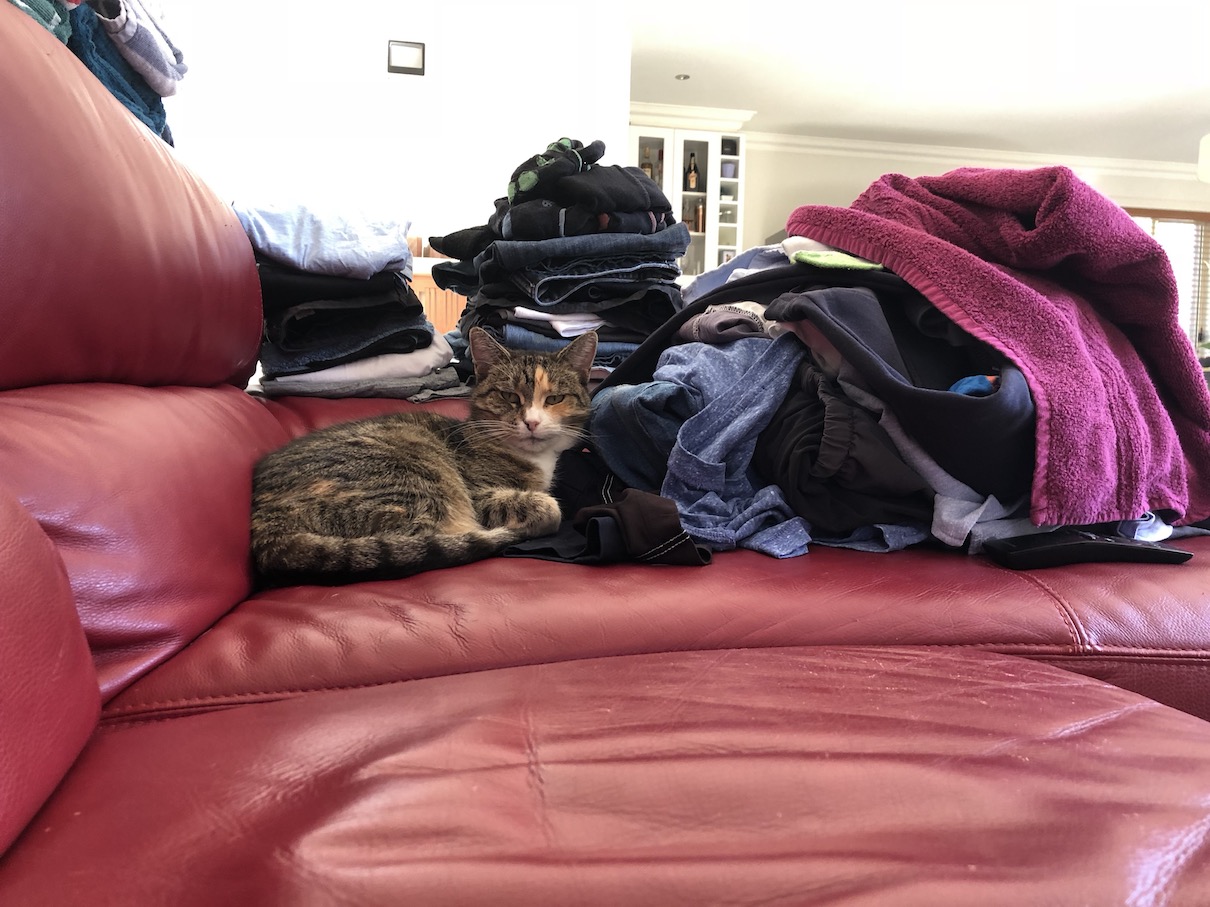}  
    \caption{}
    \label{fig:sample-images:cat}
  \end{subfigure}
  \hfill
  \begin{subfigure}[b]{0.49\linewidth}
    \includegraphics[width=\linewidth]{images/000000196009} 
    \caption{}
    \label{fig:sample-images:carrot}
  \end{subfigure}
  \caption{
    \textit{Left:} The top labels for each service do not intersect, with each having a varied ontology:~$T_{i}$~=~\{~\googleapi{} = \{`black'\}, \azureapi{} = \{`indoor'\}, \awsapi{} = \{`slide', `toy'\}~\}. (Service \awsapi{} returns \textit{both} `slide' and `toy' with equal confidence.)
    \textit{Right:} The top labels for each service focus on disparate subjects in the image: $T_{i} = $~\{~\googleapi{}~=~\{ `carrot' \}, \azureapi{} = \{ `indoor'~\}, \awsapi{} = \{ `spoon' \} \}.
    }
\end{figure}

Should she decide if the performance of a particular service isn't to her needs, then the vocabulary used for these labels becomes inconsistent for all other images; that is, the top label sets per service for \cref{fig:sample-images:cat} shows no intersection at all. Furthermore, the part of the image each service focuses on may not be consistent for their top labels; in \cref{fig:sample-images:carrot}, service \googleapi{}'s top label focuses on the vegetable (`carrot'), service \awsapi{} focuses on the `spoon', while service \azureapi{}'s focus is that the image is `indoor's. It is interesting to note that service \azureapi{} focuses on the scene matter (indoors) rather than the subject matter. (Furthermore, we do not actually know if the image in \cref{fig:sample-images:carrot} was taken indoors.)

Hence, developers should ensure that the vocabulary used by a particular service is right for them before implementation. As each service does not work to the same standardised model, trained with disparate training data, and tuned differently, results will differ despite the same input. This is unlike deterministic systems: for example, switching from AWS Object Storage to Google Cloud Object storage will conceptually provide the same output (storing files) for the same input (uploading files). However, computer vision services do not agree on the top label for images, and therefore developers are likely to be vendor locked, making changes between services non-trivial.

\subsubsection{Semantic consistency where $|T_{i}| > 1$}

\begin{figure}[t]
  \centering
  \begin{subfigure}[b]{0.49\linewidth}
    \includegraphics[width=\linewidth]{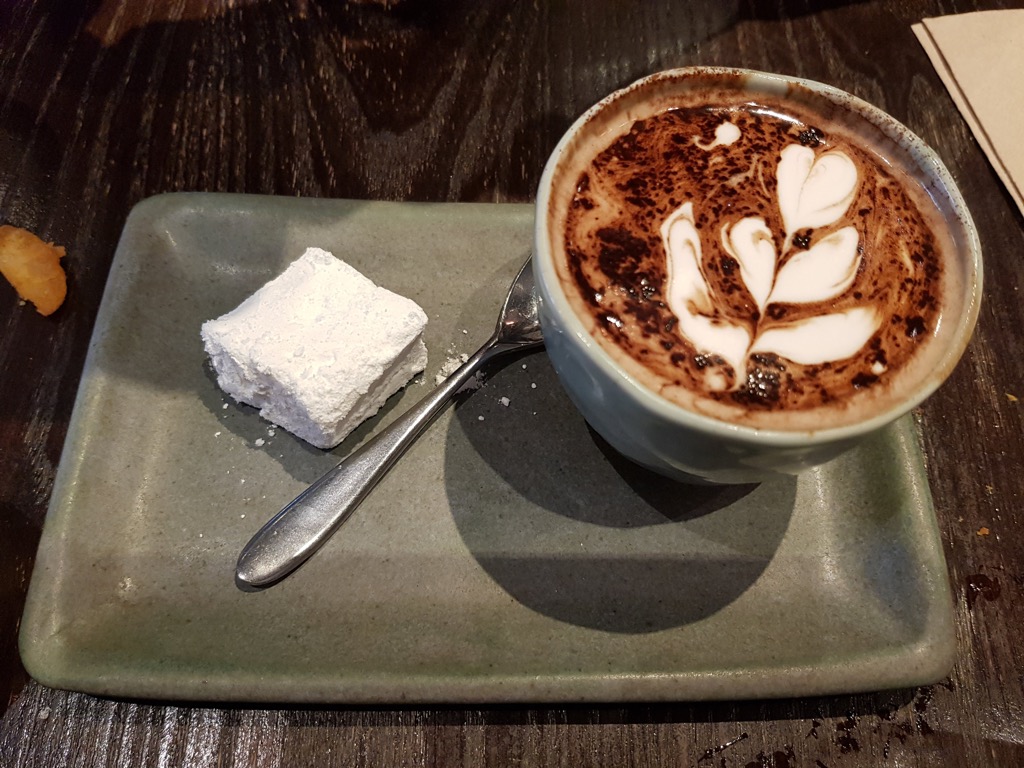}  
    \caption{}
    \label{fig:sample-images:coffee}
  \end{subfigure}
  \hfill
  \begin{subfigure}[b]{0.49\linewidth}
    \includegraphics[width=\linewidth]{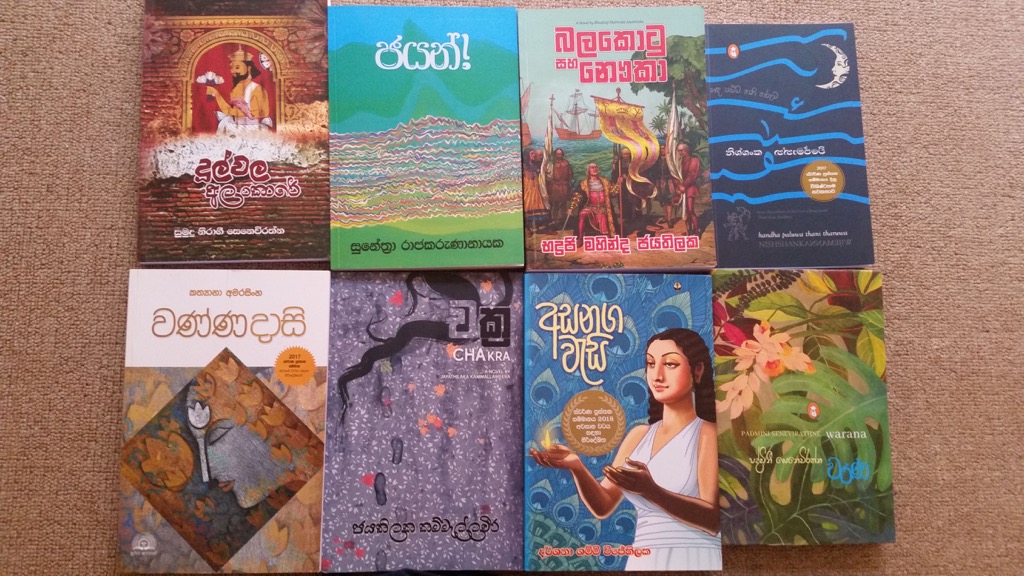} 
    \caption{}
    \label{fig:sample-images:books}
  \end{subfigure}
  \caption{
    \textit{Left:} Service \awsapi{} is 98.49\% confident of the following labels:~\{~`beverage', `chocolate', `cup', `dessert', `drink', `food', `hot chocolate'~\}. However, it is up to the developer to decide which label to persist with as all are returned.
    \textit{Right:} Service \azureapi{} persistently returns a top label set of \{~`book', `several'~\}. Both are semantically correct for the image, but disparate in what the label is to describe.
    }
\end{figure}

Service \awsapi{} returns two top labels for \cref{fig:sample-images:cat}; `slide' and `toy'. More than one top label is typically returned in service \awsapi{} (80.00\%, 56.97\%, and 81.66\% of all images for all three data sets, respectively) though this also occurs in \azureapi{} in the large (4.97\% of all images) and COCOVal17 data sets (2.38\%). Semantic inconsistencies of what this label conceptually represents becomes a concern as these labels have confidences of \textit{equal highest} consistency. Thus, some services are inconsistent in themselves and cannot give a guaranteed answer of what exists in an image; services \awsapi{} and \azureapi{} have multiple top labels, but the respective services cannot `agree' on what the top label actually is. In \cref{fig:sample-images:coffee}, service \awsapi{} presents a reasonably high confidence for the set of 7 top labels it returns, however there is too much diversity ranging from a `hot chocolate' to the hypernym `food'. Both are technically correct, but it is up to the developer to decide the level of hypernymy to label the image as. We also observe a similar effect in \cref{fig:sample-images:books}, where the image is labelled with both the subject matter and the number of subjects per image.

Thus, a taxonomy of ontologies is unknown; if a `border collie' is detected in an image, does this imply the hypernym `dog' is detected, and then `mammal', then `animal', then `object'? Only service \azureapi{} documents a taxonomy for capturing what level of scope is desired, providing what it calls the `86-category' concept as found in its how-to guide:

\begin{quote}
  \itshape
  ``Identify and categorize an entire image, using a category taxonomy with parent/child hereditary hierarchies. Categories can be used alone, or with our new tagging models.'' 
  \upshape\citepweb{Azure:WhatIs}
\end{quote}

 Thus, even if Rosa implemented conceptual similarity analysis for the image, the top label set may not provide sufficient information to derive at a conclusive answer, and if simply relying on only one label in this set, information such as the duplicity of objects (e.g., `several' in \cref{fig:sample-images:books}) may be missed.

\begin{figure}[t!]
  \centering
  \includegraphics[width=0.96\linewidth]{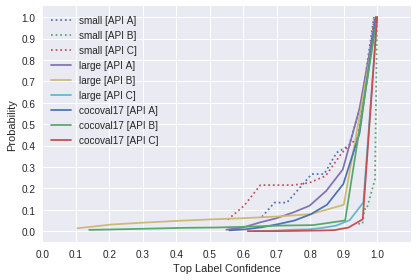}
  \caption{Cumulative distribution of the top labels' confidences. One in nine images return a top label(s) confident to $\gtrapprox97\%$, though there is a wider distribution for service \googleapi{}.}  
  \label{fig:topconfidence-cdf}
\end{figure}

\subsection{Consistency of confidence}

\subsubsection{Consistency of top label's confidence}

In \cref{fig:topconfidence-cdf}, we see that there is high probability that top labels have high confidences for all services. In summary, one in nine images uploaded to any service will return a top label confident to at least 97\%. However, there is higher probability for service \googleapi{} returning a lower confidence, followed by \azureapi{}. The best performing service is \awsapi{}, with 90\% of requests having a top label confident to~$\gtrapprox95\%$, when compared to $\gtrapprox87\%$ and $\gtrapprox93\%$ for services \googleapi{} and \azureapi{}, respectively.

Therefore, Rosa could generally expect that the top labels she receives in her images do have high confidence. That is, each service will return a top label that they are confident about. This result is expected, considering that the `top' label is measured by the highest confidence, though it is interesting to note that some services are generally more confident than others in what they present back to users.

\begin{table*}[t!]
\renewcommand{\arraystretch}{1.3}
\caption{Ratio of the top labels (to images) that remained the top label but changed confidence values between intervals.}
\label{tab:delta-confidences}
\centering
\begin{tabular}{|c||c|c|c||c|c|c|c|}
  \hline
  \textbf{Service} & \textbf{Small} & \textbf{Large} & \textbf{COCOVal17} & \textbf{$\mu(\delta_{c})$} & \textbf{$\sigma(\delta_{c})$} & Median$(\delta_{c})$ & Range$(\delta_{c})$ \\
  \hline
  \hline
  \googleapi{} & 53.33\%  & 59.19\%  & 44.92\% & \num{9.62E-08} & \num{6.84E-08} & \num{5.96E-08} & $[\num{5.96E-08}, \num{6.56E-07}]$  \\
  \hline
  \azureapi{}  & 0.00\%  & 0.00\%    & 0.02\%  & - & - & - & -  \\
  \hline
  \awsapi{}    & 33.33\%  & 41.36\%  & 15.60\% &  \num{5.35E-07} & \num{8.76E-07} & \num{3.05E-07} & $[\num{1.27E-07}, \num{1.13E-05}]$  \\
  \hline
\end{tabular}
\end{table*}

\begin{figure}[t]
  \centering
  \includegraphics[width=0.96\linewidth]{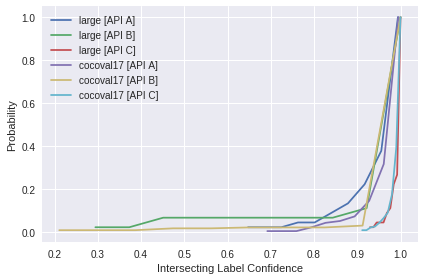}
  \caption{Cumulative distribution of intersecting labels top labels' confidences. The small data set is intentionally removed due to low intersections of labels (see \cref{tab:intersect-of-labels}).}
  \label{fig:intersecting-cdf}
\end{figure}

\subsubsection{Consistency of intersecting top label's confidence}

Even where all three services do agree on a set of top labels, the disparity of how much they agree by is still of importance. Just because three services agree that an image contains consistent top labels, they do not always have a small spread of confidence. In \cref{fig:sample-images:cake}, the three services agree with $\sigma~=~0.277$, significantly larger than that of all images in general $\sigma~=~0.0831$. \Cref{fig:intersecting-cdf} displays the cumulative distribution of all intersecting top labels' confidence values, presenting slightly similar results to that of \cref{fig:topconfidence-cdf}.

\begin{figure}[t]
  \centering
  \includegraphics[width=0.55\linewidth]{images/000000095707}    
  \caption{All three services agree the top label for the above image is `food', but the confidences to which they agree by vary significantly. Service \awsapi{} is most confident to 94.93\% (in addition with the label `bread'); service \googleapi{} is the second most confident to 84.32\%; service \azureapi{} is the least confident with 41.39\%.}
  \label{fig:sample-images:cake}
\end{figure}

\subsection{Evolution risk}

\subsubsection{Label Stability}

Generally, the top label(s) did not evolve in the evaluation period. 16.19\% and 5.85\% of images did change their top label(s) in the Large and COCOVal17 data sets in service \googleapi{}. Thus, top labels are stable but not guaranteed to be constant.

\subsubsection{Confidence Stability}

Similarly, where the top label(s) remained the same from one interval to the next, the confidence values were stable. \Cref{tab:delta-confidences} displays the proportion of images that changed their top label's confidence values with various statistics on the confidence deltas between snapshots ($\delta_{c}$). However, this delta is so minuscule that we attribute such changes to statistical noise.

\section{Recommendations}
\label{sec:recommendations}

\subsection{Recommendations for intelligent service users}

\subsubsection{Test with a representative ontology for the particular use case}
Rosa should ensure that in her testing strategies for the app she develops, there is an ontology focus for the types of vocabulary that are returned. Additionally, we noted that there was a sudden change in case for services \googleapi{} and \awsapi{}; for all comparative purposes of labels, each label should be lower-cased. 

\subsubsection{Incorporate a specialised intelligent service testing methodology into the development lifecycle}
Rosa can utilise the different aspects of consistency as outlined in this paper as part of her quality strategy. To ensure results are correct over time, we recommend developers create a representative data set of the intended application's data set and evaluate these changes against their chosen service frequently. This will help identify when changes, if any, have occurred if vendors do not provide a line of communication when this occurs.

\subsubsection{Intelligent services are not `plug-and-play'}
Rosa will be locked into whichever vendor she chooses as there is inherent inconsistency between these services in both the vocabulary and ontologies that they use. We have demonstrated that very few services overlap in their vocabularies, chiefly because they are still in early development and there is yet to be an established, standardised vocabulary that can be shared amongst the different vendors.
 Issues such as those shown in \cref{ssec:findings:consistency-of-labels} can therefore be avoided.
 
Throughout this work, we observed that the terminologies used by the various vendors are different. Documentation was studied, and we note that there is inconsistency between the ways techniques are described to users. We note the disparity between the terms `detection', `recognition', `localisation' and `analysis'. This applies chiefly to object- and facial-related techniques. Detection applies to facial detection, which gives bounding box coordinates around all faces in an image. Similarly, localisation applies the same methodology to disparate objects in an image and labels them. In the context of facial `recognition', this term implies that a face is \textit{recognised} against a known set of faces. Lastly, `analysis' applies in the context of facial analysis (gender, eye colour, expression etc.); there does not exist a similar analysis technique on objects.

We notice similar patterns with object `tagging', `detection' and `labelling'. Service \googleapi{} uses `Entity Detection' for object categorisation, service \azureapi{} uses `Image Tagging', and service \awsapi{} uses the term `Detect Labels' : conceptually, these provide the same functionality but the lack of consistency used between all three providers is concerning and leaves room for confusion with developers during any comparative analyses. Rosa may find that she wants to label her images into day/night scenes, but this in turn means the `labelling' of varying objects. There is therefore no consistent standards to use the same terminology for the same concepts, as there are in other developer areas (such as Web Development).

\subsubsection{Avoid use in safety-critical systems}
 We have demonstrated in this paper that both labels and confidences are stable but not constant; there is still an evolution risk posed to developers that may cause unknown consequences in applications dependent on these computer vision services. Developers should avoid their use in safety critical systems due to the lack of visible changes.
 
\subsection{Recommendations for intelligent service providers}

\subsubsection{Improve the documentation}
Rosa does not know that service \googleapi{} returns back `carrot' for its top response, with service \awsapi{} returning `spoon' (\cref{fig:sample-images:carrot}). She is unable to tell the service's API where to focus on the image. Moreover, how can she toggle the level of specificity in her results? She is frustrated that service \awsapi{} can detect `chocolate', `food' and also `beverage' all as the same top label in \cref{fig:sample-images:coffee}: what label is she to choose when the service is meant to do so for her, and how does she get around this? Thus, we recommend vendors to improve the documentation of services by making known the boundary set of the training data used for the algorithms. By making such information publicly available, developers would be able to review the service's specificity for their intended use case (e.g., maybe Rosa is satisfied her app can catalogue `food' together, and in fact does not want specific types of foods (`hot chocolate') catalogued). We also recommend that vendors publish usage guidelines should that include details of priors and how to evaluate the specific service results.

Furthermore, we did not observe that the vendors documented how some images may respond with multiple labels of the exact same confidence value. It is not clear from the documentation that response objects can have duplicate top values, and tutorials and examples provided by the vendors do not consider this possibility. It is therefore left to the developer to decide which label from this top set of labels best suits for their particular use case; the documentation should describe that a rule engine may need to be added in the developer's application to verify responses. The implications this would have on maintenance would be significant.

\subsubsection{Improve versioning}
We recommend introducing a versioning system so that a model can be used from a specific date in production systems: when Rosa tests her app today, she would like the service to remain \textit{static} the same for when her app is deployed in production tomorrow. Thus, in a request made to the vendor, Rosa could specify what date she ran her app's QA testing on so that she knows that henceforth these model changes will not affect her app.

\subsubsection{Improve Metadata in Response}
Much of the information in these services is reduced to a single confidence value within the response object, and the details about training data and the internal AI architecture remains unknown; little metadata is provided back to developers that encompass such detail. Early work into model cards and datasheets~\citep{Mitchell:2018in,Gebru:2018wh} suggests more can be done to document attributes about ML systems, however at a minimum from our work, we recommend including a reference point via the form of an additional identifier. This identifier must also permit the developers to submit the identifier to another API endpoint should the developer wish to find further characteristics about the AI empowering the intelligent service, reinforcing the need for those presented in model cards and datasheets. For example, if Rosa sends this identifier she receives in the response object to the intelligent service descriptor API, she could find out additional information such as the version number or date when the model was trained, thereby resolving potential evolution risk, and/or the ontology of labels.
 
\subsubsection{Apply constraints for predictions on all inputs}
In this study, we used some images with intentionally disparate, and noisy objects. If services are not fully confident in the responses they  give back, a form of customised error message should be returned. For example, if Rosa uploads an image of 10 various objects on a table, rather than returning a list of top labels with varying confidences, it may be best to return a `too many objects' exception. Similarly, if Rosa uploads a photo that the model has had no priors on, it might be useful to return an `unknown object' exception than to return a label it has no confidence of. We do however acknowledge that current state of the art computer vision techniques may have limits in what they can and cannot detect, but this limitation can be exposed in the documentation to the developers.

A further example is sending a one pixel image to the service, analogous to sending an empty file. When we uploaded a single pixel white image to service \googleapi{}, we received responses such as `microwave oven', `text', `sky', `white' and `black' with confidences ranging from 51--95\%. Prior checks should be performed on all input data, returning an `insufficient information' error where any input data is below the information of its training data.
 
\section{Threats to Validity}
\label{sec:limitations}

\subsection{Internal Validity}
Not all computer vision services were assessed. As suggested in \cref{sec:method}, we note that there are other computer vision services such as IBM Watson. Many services from Asia were also not considered due to language barriers (of the authors) in assessing these services. We limited our study to the most popular three providers (outside of Asia) to maintain focus in this body of work.
 
A custom confidence threshold was not set. All responses returned from each of the services were included for analysis; where confidences were low, they were still included for analysis. This is because we used the default thresholds of each API to hint at what real-world applications may be like when testing and evaluating these services.

The label string returned from each service was only considered. It is common for some labels to respond back that are conceptually similar (e.g., `car' vs. `automobile') or grammatically different (e.g., `clothes' vs. `clothing'). While we could have employed more conceptual comparison or grammatical fixes in this study, we chose only to compare lowercased labels and as returned. We leave semantic comparison open to future work.

Only introductory analysis has been applied in assessing the documentation of these services. Further detailed analysis of documentation quality against a rigorous documentation quality framework would be needed to fortify our analysis of the evolution of these services' documentation.

\subsection{External Validity}

The documentation and services do change over time and evolve, with many allowing for contributions from the developer community via GitHub. We note that our evaluation of the documentation was conducted on a single date (see \cref{sec:method}) and acknowledge that the documentation may have changed from the evaluation date to the time of this publication. We also acknowledge that the responses and labelling may have evolved too since the evaluation period described and the date of this publication. Thus, this may have an impact on the results we have produced in this paper compared to current, real-world results. To mitigate this, we have supplied the raw responses available online~\citepweb{Results:RawData}.

Moreover, in this paper we have investigated \textit{computer vision} services. Thus, the significance of our results to other domains such as natural language processing or audio transcription is, therefore, unknown. Future studies may wish to repeat our methodology on other domains to validate if similar patterns occur; we remain this open for future work.

\subsection{Construct Validity}

It is not clear if all the recommendations proposed in \cref{sec:recommendations}  are feasible or implementable in practice. Construct validity defines how well an experiment measures up to its claims; the experiments proposed in this paper support our three hypotheses but these have been conducted in a clinical condition. Real-world case studies and feedback from developers and providers in industry would remove the controlled nature of our work.

\section{Conclusions \& Future Work}
\label{sec:conclusions}

This study explored three popular computer vision services over an 11 month longitudinal experiment to determine if these services pose any evolution risk or inconsistency. We find that these services are generally stable but behave inconsistently; responses from these services do change with time and this is not visible to the developers who use them. Furthermore, the limitations of these systems are not properly conveyed by vendors. From our analysis, we present a set of recommendations for both intelligent service vendors and developers.

Standardised software quality models (e.g.,~\citep{ISO9126:1999}) target maintainability and reliability as primary characteristics. Quality software is stable, testable, fault tolerant, easy to change and mature. These computer vision services are, however, in a nascent stage, difficult to evaluate, and currently are not easily interchangeable. Effectively, the intelligent service response objects are shifting in material ways to developers, albeit slowly, and vendors do not  communicate this evolution or modify API endpoints; the endpoint remains static but the content returned does not despite the same input.

There are many potential directions stemming from this work. 
To start, we plan to focus on preparing a more comprehensive datasheet specifically targeted at what should be documented to application developers, and not data scientists. 
Reapplying this work in real-world contexts, that is, to get real developer opinions and study production grade systems, would also be beneficial to understand these  phenomena in-context. This will help us clarify if such changes are a real concern for developers (i.e., if they really need to change between services, or the service evolution has real impact on their applications).
We also wish to refine and systematise the method used in this study and develop change detectors that can be used to identify evolution in these services that can be applied to specific ML domains (i.e., not just computer vision), data sets, and API endpoints, thereby assisting application developers in their testing strategies.
Moreover, future studies may wish to expand the methodology applied by refining how the responses are compared. As there does not yet exist a standardised list of terms available between services, labels could be \textit{semantically} compared instead of using exact matches (e.g., by using stem words and synonyms to compare similar meanings of these labels), similar to previous studies \citep{Ohtake:2019vi}.

This paper has highlighted only some high-level issues that may be involved in using these evolving services. The laws of software evolution suggest that for software to be useful, it must evolve \citep{THOMAS2014457,1572302}. There is, therefore, a trade-off, as we have shown, between consistency and evolution in this space. For a component to be stable, any changes to dependencies it relies on must be communicated. We are yet to see this maturity of communication from intelligent service providers. Thus, developers must be cautious between integrating intelligent components into their applications at the expense of stability; as the field of AI is moving quickly, we are more likely to see further instability and evolution in intelligent services as a consequence.

\balance
\bibliography{papers,pc}
\bibliographystyleweb{IEEEtranN}
\bibliographyweb{web}

\end{document}